\documentclass[showpacs,twocolumn,amssymb]{aastex62}

\usepackage{graphicx}
\usepackage{dcolumn}
\usepackage{bm}
\usepackage{amsmath}
\usepackage{amssymb}

\newcommand{\g}{\text{ g}}
\newcommand{\K}{\text{ K}}

\newcommand{\erg}{\text{ erg}}
\newcommand{\cm}{\text{ cm}}

\newcommand{\mpc}{\text{ Mpc}}
\newcommand{\pc}{\text{ pc}}
\newcommand{\second}{\text{ s}}
\newcommand{\Hz}{\text{ Hz}}
\newcommand{\yr}{\text{ yr}}
\newcommand{\ud}{\mathrm{d}}

\newcommand{\cqg}{\textit{Class. Quantum Grav.}}
\newcommand{\livingrev}{\textit{Living Rev Relativ}}
\newcommand{\sci}{\textit{Science}}

\begin{document}

\title{A bright electromagnetic counterpart to extreme mass ratio inspirals}

\author[0000-0002-3822-0389]{Y. Y. Wang}
\affiliation{School of Astronomy and Space Science, Nanjing University, Nanjing 210093, China}

\author[0000-0003-4157-7714]{F. Y. Wang}
\affiliation{School of Astronomy and Space Science, Nanjing University, Nanjing 210093, China}
\affiliation{Key Laboratory of Modern Astronomy and Astrophysics (Nanjing University), Ministry of Education, Nanjing 210093, China}

\author[0000-0002-5400-3261]{Y. C. Zou}
\affiliation{School of Physics, Huazhong University of Science and Technology, Wuhan 430074, China}

\author{Z. G. Dai}
\affiliation{School of Astronomy and Space Science, Nanjing University, Nanjing 210093, China}
\affiliation{Key Laboratory of Modern Astronomy and Astrophysics (Nanjing University), Ministry of Education, Nanjing 210093, China}

\correspondingauthor{F. Y. Wang}
\email{fayinwang@nju.edu.cn}

\date{\today}

\begin{abstract}
The extreme mass ratio inspiral (EMRI), defined as a
stellar-mass compact object inspiraling into a supermassive black
hole (SMBH), has been widely argued to be a low-frequency
gravitational wave (GW) source. EMRIs providing accurate measurements
of black hole mass and spin, are one of the primary interests for Laser
Interferometer Space Antenna (LISA). However, it is usually believed
that there are no electromagnetic (EM) counterparts to EMRIs. Here we
show a new formation channel of EMRIs with tidal disruption flares
as EM counterparts. In this scenario, flares can be
produced from the tidal stripping of the helium (He) envelope of a
massive star by an SMBH. The left compact core of the massive star
will evolve into an EMRI. We find that, under certain initial
eccentricity and semimajor axis, the GW frequency of the inspiral can enter
LISA band within 10 $\sim$ 20 years, which makes the tidal disruption
flare an EM precursor to EMRI. Although
the event rate is just $2\times 10^{-4}~\rm Gpc^{-3}yr^{-1}$, this
association can not only improve the localization accuracy of LISA
and help to find the host galaxy of EMRI, but also serve as
a new GW standard siren for cosmology.
\end{abstract}


\section{Introduction}
The detection of GW170817/GRB 170817A heralds the era of
gravitational-wave (GW) multimessenger astronomy
\citep{Abbott17a}. The neutron star-neutron star (NS-NS) and neutron
star-black hole (NS-BH) mergers accompanied by electromagnetic (EM)
counterparts offer a standard siren for cosmology, which can
independently constrain the Hubble constant $H_0$
\citep{Abbott17b,Chen18,Wang18} , calibrate luminosity
correlations of $\gamma$-ray bursts \citep{Wang19} and so on.
In addition to mergers of compact binaries, the
extreme mass ratio inspiral (EMRI)
\citep{Amaro-Seoane07,Gair13,Babak17}, which originates from the inspiral of a
compact object into a supermassive black hole (SMBH), is another source of
gravitational wave. Detecting
EMRIs is one of the most crucial scientific goals of future
space-based GW detectors such as Laser Interferometer Space Antenna
(LISA)
\citep{Danzmann00,Phinney02,Amaro-Seoane17,Babak17,Amaro-Seoane18},
TianQin Project \citep{Luo15} and Taiji Program \citep{Hu2017}.
Nevertheless, LISA can only determine the sky location and
luminosity distance of EMRI to a few square degrees
\citep{Cutler98} and $10\%$ precision \citep{Babak17} respectively,
which may not identify the host galaxy uniquely. In this
case, statistical methods ought to be used to determine the
host galaxy. However, the redshift obtained in this way is not
independent of the luminosity distance \citep{Amaro-Seoane07}.
On the contrary, the EMRIs, if having EM counterparts, will
serve as a powerful standard siren. However, it
seems that there is no EM signal accompanying EMRIs
\citep{Amaro-Seoane07}, which poses the main obstacle for
cosmological application.

The ``standard" formation channel of EMRIs is the capture of
a compact object (white dwarf (WD), NS or BH) by an SMBH
\citep{Sigurdsson97,Amaro-Seoane07}. Other processes include tidal
separation of compact binaries, formation or capture of massive
stars in accretion discs and so on
\citep{Amaro-Seoane07,Maggiore18}.

In this paper, we exploit a new formation channel for EMRIs. In our model, the EMRI signal comes from the inspiral of
a massive star which was tidally stripped by an SMBH.
Our paper is organized as follows. In Section \ref{Sec:2},
we describe the tidal stripping of massive star's envelopes. The structure and orbital evolution of the remnant core are
introduced in Section \ref{Sec:3}. The signal-to-noise ratio of the
EMRI is estimated in Section \ref{Sec:5}. A discussion of the EMRI
rate and a brief summary are given in Section \ref{Sec:6} and
\ref{Sec:7}, respectively.

\begin{figure*}
    \centering
    \includegraphics[width=0.5\textwidth]{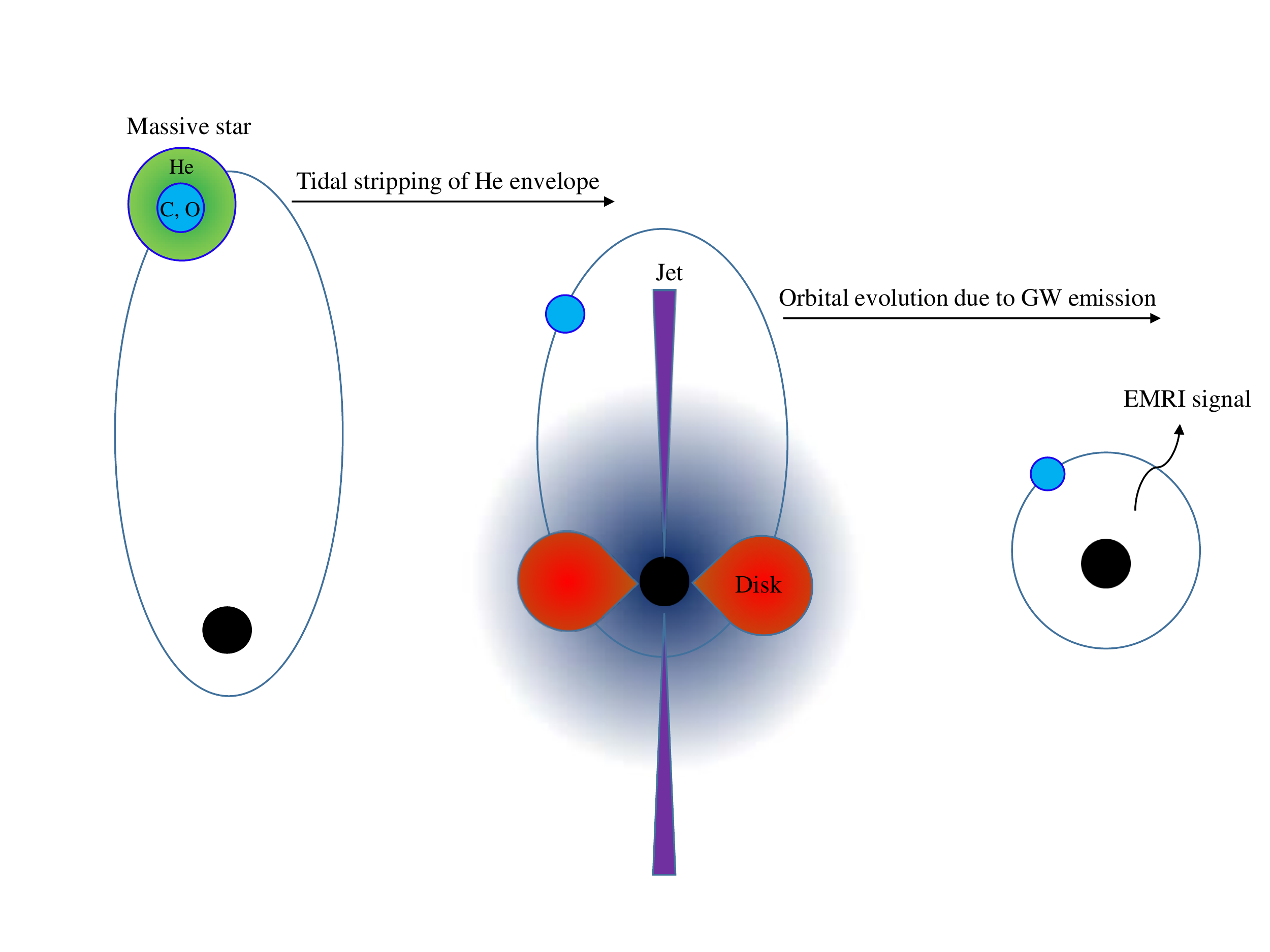}
    \caption{A schematic diagram of the mechanism for observing EMRI and relevant tidal disruption EM signal in our model.
    Initially, an onion-skin layered massive star orbits an SMBH. After the He envelope gets tidally stripped, X-ray flares
    are produced by the accretion flow. In some cases, a relativistic jet could be
    launched by the accreting SMBH. Eventually, the left compact C-O core with density about $10^6\g\cm^{-3}$
    will inspiral into the SMBH and produce EMRI signal in the LISA band.}
    \label{Fig:Schematic}
\end{figure*}

\section{Tidal stripping of stellar envelope and flares}
\label{Sec:2} When a star passes close enough through an
SMBH, it will be torn apart by the tidal force
\citep{Hills75,Rees88,Evans89,Phinney89}. A star with density $\rho$
is tidally disrupted when the work exerted over it by the tidal
force exceeds its binding energy \citep{Rees88,Amaro-Seoane18}. The
tidal radius can be calculated from
\begin{equation}
R_{\text{T}} \simeq R_*\bigg(\frac{M_{\text{BH}}}{M_*}\bigg)^{1/3} =
\bigg(\frac{3M_{\text{BH}}}{4\pi \rho}\bigg)^{1/3},
\end{equation}
where $M_{\text{BH}}$ is the mass of the BH, $R_*$ and $M_*$ are the
stellar radius and stellar mass respectively. The penetration factor $\beta$ defines
the strength of the tidal interaction exerted on the star
\citep{Carter82}
\begin{equation}
\beta = \frac{R_T}{R_p},
\end{equation}
where $R_p = a(1-e)$ is the pericenter.

Besides the whole star, the envelopes of evolving stars can
also be tidally stripped. For example, the ultraviolet-optical
transient PS1-10jh can be explained by tidal disruption of a
helium-rich stellar core, which is considered the remnant
of a tidally stripped red giant (RG) star \citep{Gezari12}.
Furthermore, \citet{Bogdanovic14} studied the tidal stripping
of an RG star's envelope by an SMBH and the subsequent inspiral of the core toward
the BH. Typically, a massive star has a so-called
``onion-skin" structure at the end of its evolution, where each
shell has different chemical compositions and mass densities
\citep{Woosley02}. The outer layers have much lower
densities than the core, which makes them more vulnerable
to tidal forces. Therefore, a massive star may lose its envelopes partially or completely when it passes close enough
through an SMBH,
leaving a dense core on a highly eccentric orbit
\citep{DiStefano01,Kobayashi04,Davies05,Amaro-Seoane07,Guillochon13}.
The tidal disruption flares of main sequence stars and helium stars
accompanied by GW bursts were investigated previously
\citep{Kobayashi04}. However, this type of GW bursts can not be
observed if luminosity distance $d_L$ is larger than $20\mpc$
\citep{Kobayashi04}, which limits its cosmological applications.

Here, we propose a new formation channel for EMRIs with EM
precursors. In our model, we assume that a massive star has lost
H envelope during the red supergiant period. It is in
the He burning stage \citep{Heger03} and the densities of different
layers vary from $10^{0-3}\g\cm^{-3}$ (He envelope) to
$10^{5-6}\g\cm^{-3}$ (carbon-oxygen (C-O) core) \citep{Woosley02}.
After the He envelope is tidally stripped by the SMBH and the
C-O core finally inspirals into the SMBH, we can detect a tidal
disruption event (TDE) and the subsequent EMRI signal. For a TDE, we can
identify its host galaxy and determine the redshift through
spectral lines observation. With 
the luminosity distance $d_L$ from the EMRI signal and the redshift $z$, we have a new
type of standard siren. The luminosity distance can
be determined to $10\%$ precision at $z = 1$. Figure
\ref{Fig:Schematic} shows a schematic picture of our model.

Since the typical density of He envelope is
$10^3\g\cm^{-3}$ \citep{Woosley02}, the tidal stripping
should take place in an orbit of semimajor axis $a\sim$ a
few $10^{-6}\pc$ and eccentricity $e=0.90\sim 0.98$.
For our scenario to work successfully, the
tidal radius of He envelope $R_{\text{T, He}}$ should be larger than
the innermost stable circular orbit (ISCO) radius $R_{\text{ISCO}}$.
Meanwhile, the
tidal radius of C-O core $R_{\text{T,C-O}}$ should be smaller than
$R_{\text{ISCO}}$. Therefore, the feasible mass range of centeral SMBH is approximately $3\times 10^4\sim 8\times 10^5 M_\odot$.

Below we show the observational properties
of the tidal disruption flare in our model. The energy required to strip the stellar envelope
is \citep{Davies05}
\begin{equation}
E_{\text{strip}}\sim \frac{G M_c M_e}{R_c},
\end{equation}
where $M_c$, $R_c$ and $M_e$ are core mass, core radius and stripped
envelope mass respectively.
If the tidal disruption happens on a highly eccentric orbit, about
half of the debris will fall back to the BH, in which case
the luminosity of TDE is supposed to follow the standard
$t^{-5/3}$ decay rate  \citep{Rees88,Evans89,Phinney89}.
For a 15$M_{\odot}$ star, the masses of the core
and stellar debris are about 3 $M_\odot$ and 1$M_\odot$
respectively. Assuming $f$ is the fraction of the accreted stellar
envelope relative to the massive star, then the bound material
returns to pericenter at a rate
\begin{equation}
\begin{aligned}
\dot{M} &\simeq \frac{1}{3}\frac{f M_*}{P_{\text{min}}} \bigg(\frac{t}{P_{\text{min}}}\bigg)^{-5/3}\\
&\simeq 6\times 10^2 M_\odot\yr^{-1}\bigg(\frac{f}{0.25}\bigg)\\
&\cdot
\bigg(\frac{R_*}{10^{10}\cm}\bigg)^{-3/2} \bigg(\frac{M_*}{4
    M_\odot}\bigg)^2 \bigg(\frac{M_{\rm BH}}{10^5
    M_\odot}\bigg)^{-1/2}\bigg(\frac{t}{P_{\text{min}}}\bigg)^{-5/3},
\end{aligned}
\end{equation}
where
\begin{equation}
\begin{aligned}
P_{\text{min}} &= \frac{2\pi R_p^3}{(GM)^{1/2}(2R_*)^{3/2}} \\
&\simeq 6\times 10^3\second \bigg(\frac{R_*}{10^{10}\cm}\bigg)^{3/2}
\bigg(\frac{M_*}{4 M_{\odot}}\bigg)^{-1} M_5^{1/2},
\end{aligned}
\end{equation}
is the shortest Keplerian orbital period
\citep{Ulmer99,Bogdanovic14}; $M_5$ is defined as $M_5 \equiv M_{\text{BH}}/(10^5
M_{\odot})$.

The luminosity of the accretion flow falling back to the SMBH
is \citep{Bogdanovic14}
\begin{equation}
\begin{aligned}
L &= \epsilon\dot{M}c^2 \\
&\simeq 2\times 10^{48}\erg\second^{-1} \bigg(\frac{f}{0.25}\bigg) \bigg(\frac{\epsilon}{0.057}\bigg) \bigg(\frac{R_*}{10^{10}\cm}\bigg)^{-3/2} \\
& \cdot\bigg(\frac{M_*}{4 M_{\odot}}\bigg)^2 M_5^{-1/2}
\bigg(\frac{t}{P_{\text{min}}}\bigg)^{-5/3},
\end{aligned}
\end{equation}
where $\epsilon = 1-(r-2)/[r(r-3)]^{1/2}$ is the radiative
efficiency for a Schwarzschild black hole and $r$ is the orbital
radius of the debris in units of $R_g$ \citep{Bogdanovic14}.
The luminosity can be significantly larger than the Eddington limit for a
period of weeks to years \citep{Strubbe09}. When $e\le
e_{\text{crit}} = 1-2q^{-1/3}/\beta$ where $\beta\equiv M_{\text{BH}}/M_*$, the event is categorized as
eccentric TDE \citep{Hayasaki18} and all of the debris will remain
gravitationally bound to the SMBH. In these cases, the mass fallback
rate is flatter and slightly higher than the standard rate
\citep{Hayasaki18}. Besides, the fallback rate and TDE light curve
of more centrally concentrated stars show a
significant deviation from the $t^{-5/3}$ decay rate
\citep{Lodato09,Hayasaki13,Dai13,Bogdanovic14}.

The spectra of tidal flares are very complicated, which are
a superposition of blackbody spectrum and many emission lines
\citep{Strubbe09}. The temperature of the debris is
\begin{equation}
    T_{\text{eff}} = \bigg(\frac{L}{16\pi R_T^2\sigma}\bigg)^{1/4}\simeq 2\times 10^5\K \,M_5^{1/12}.
\end{equation}
The luminosity $L$ of the X-ray flares from accretion flow falling
back to the SMBH is about $10^{48}\erg\second^{-1}$ as estimated above. For
Einstein Probe under construction, which will have a field of view of 3,600 square degrees, the
flux sensitivity can be up to $10^{-10}$ erg~cm$^{-2}$~s$^{-1}$
\citep{Yuan15}. Hence, Einstein Probe can detect the X-ray flares at $z\ge
1.0$. In some cases, a TDE is accompanied by a
relativistic jet, which has been observed in the transient Swift J1644+57
\citep{Bloom11,Burrows11,Zauderer11}. If the jet points
to us, its luminosity will be much higher than that of the accretion flow.

\section{Structure and orbital evolution of the compact core}
\label{Sec:3}
\subsection{Radius expansion after tidal stripping}
After the He envelope is stripped, the core has to adjust to
a new equilibrium by expanding its radius. For solar-type stars, the
core expansion had been extensively discussed using the
mass-radius relation for the adiabatic evolution of a nested
polytrope \citep{Hjellming87,MacLeod13,Bogdanovic14}. However,
\cite{MacLeod13} showed that the assumptions for the mass-radius
relation are incorrect. Therefore, we perform a rough estimation of
the new radius using hydrostatic equilibrium equation, the first law
of thermodynamics and the relation between pressure and internal
energy density instead. The 15 $M_{\odot}$ star’s model of Woosley
\& Heger (https://2sn.org/stellarevolution/) is used to estimate the
pressure in the out layer of the C-O core before expansion.
According to \cite{Pols11}, the ideal gas assumption is taken for the
He envelope and the C-O core. We find that the core's radius will
increase just 16\%, which may not affect the tidal radius greatly.

\subsection{Requirements for EMRI formation}

In order for a compact object to become an EMRI, its orbital decay
timescale by GW emission $\tau_{\text{GW}}$ \citep{Gair06} should be
sufficiently shorter than the two-body relaxation
timescale $t_{\text{rlx}}$ \citep{Amaro-Seoane07},
\begin{equation}
\tau_{\text{GW}} < C_{\text{EMRI}}(1-e) t_{\text{rlx}}.
\end{equation}
where $C_{\text{EMRI}}$ is a numerical constant sufficiently less
than 1 and $t_{\text{rlx}}$ is about $10^9 \yr$. Otherwise, the
compact core will be deflected from its original orbit through
two-body relaxation.

\begin{figure*}
    \centering
    \includegraphics[width=0.5\textwidth]{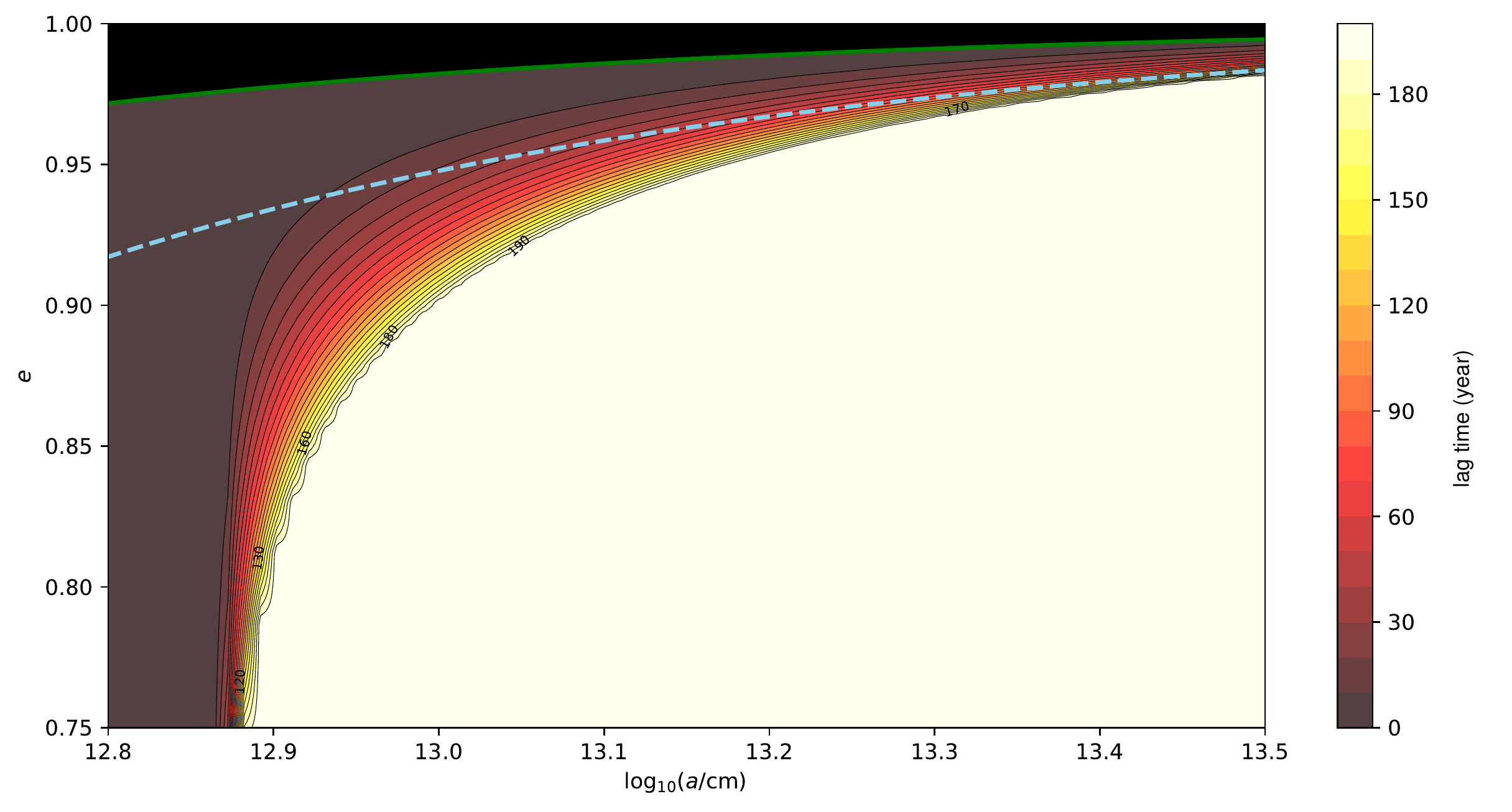}
    \caption{The contour of lag time between TDE and EMRI signal for different
    initial values of $a_0$ and $e_0$. The mass of the SMBH is chosen as $3\times 10^5 M_{\odot}$. The blue dashed line
    refers to the tidal radius $R_T$ where the He envelope (the density is taken to be $10^3 \g\cm^{-3}$)
    is disrupted. For the upper left brown region, the lag times are less than a decade, which are ideal for
    observing the TDE and EMRI association. The upper black region represents the direct plunge orbit and
    the green solid line is the upper limit for non-plunging orbit. For the right bottom region of the diagram,
    the lag times are all larger than 200 years and are not shown in detail in the contour.}
    \label{Fig:Lag time}
\end{figure*}

There is also a limitation on eccentricity $e$.
The maximal eccentricity for a non-plunging orbit
is \citep{Cutler94,Hopman05}
\begin{equation}
e_{\text{max}}(a) = -\frac{R_{\text{S}}}{2a} + \sqrt{\bigg(\frac{R_{\text{S}}}{2a}\bigg)^2-\frac{3R_{\text{S}}}{a}+1},
\end{equation}
which is depicted in Figure \ref{Fig:Lag time} with green dashed line.

\subsection{Time lag between TDE and EMRI signal}
Here, we consider the orbital evolution of the C-O core
inspiral. It is reasonable to assume that the He envelope is
completely stripped after several close encounters. Hence,
the interaction between the diffuse envelope and the core can be neglected here \citep{Amaro-Seoane07}. Furthermore, since
$a$ is only a few $10^{-6}\pc$, the encounters of the compact core
with cluster stars around the SMBH are ignored.

The semimajor axis $a$ will shrink due to
GW radiation. The Keplerian orbital evolution is given by
Peters formalism \citep{Peters64}, which is a good approximation in weak-field regime.
Apparently, there is an important factor that should be taken
seriously---the lag time between the tidal disruption and the EMRI
signal. The  EMRI enters the LISA band when its frequency $f$,
which is twice the Keplerian orbital frequency $f_{\text{orb}} =
(GM/4\pi a^3)^{1/2}$, is larger than $ 10^{-4} \Hz$. It was estimated that, for a binary
system consisting of a main sequence star and a compact object, the latter will
spend $10^2$ to $10^4$ years to spiral into
the SMBH after the main sequence star gets tidally disrupted, which
prevents the TDE from being a good precursor to the EMRI
\citep{Amaro-Seoane07}.

However, the situation can be different for a massive star.
Its tidal radius is much smaller than that of the main sequence
star, which will greatly shorten the time lag between TDE and EMRI.
The lag time contour at as a function of initial semi-major
axis $a_0$ and eccentricity $e_0$ is plotted in Figure \ref{Fig:Lag
time}. In the upper left region, the lag time is shorter than 20
years, which is ideal for observing the TDE and subsequent EMRI.

\section{Signal-to-noise ratio of EMRI}
\label{Sec:5}
\begin{figure*}
    \centering
    \includegraphics[width=0.5\textwidth]{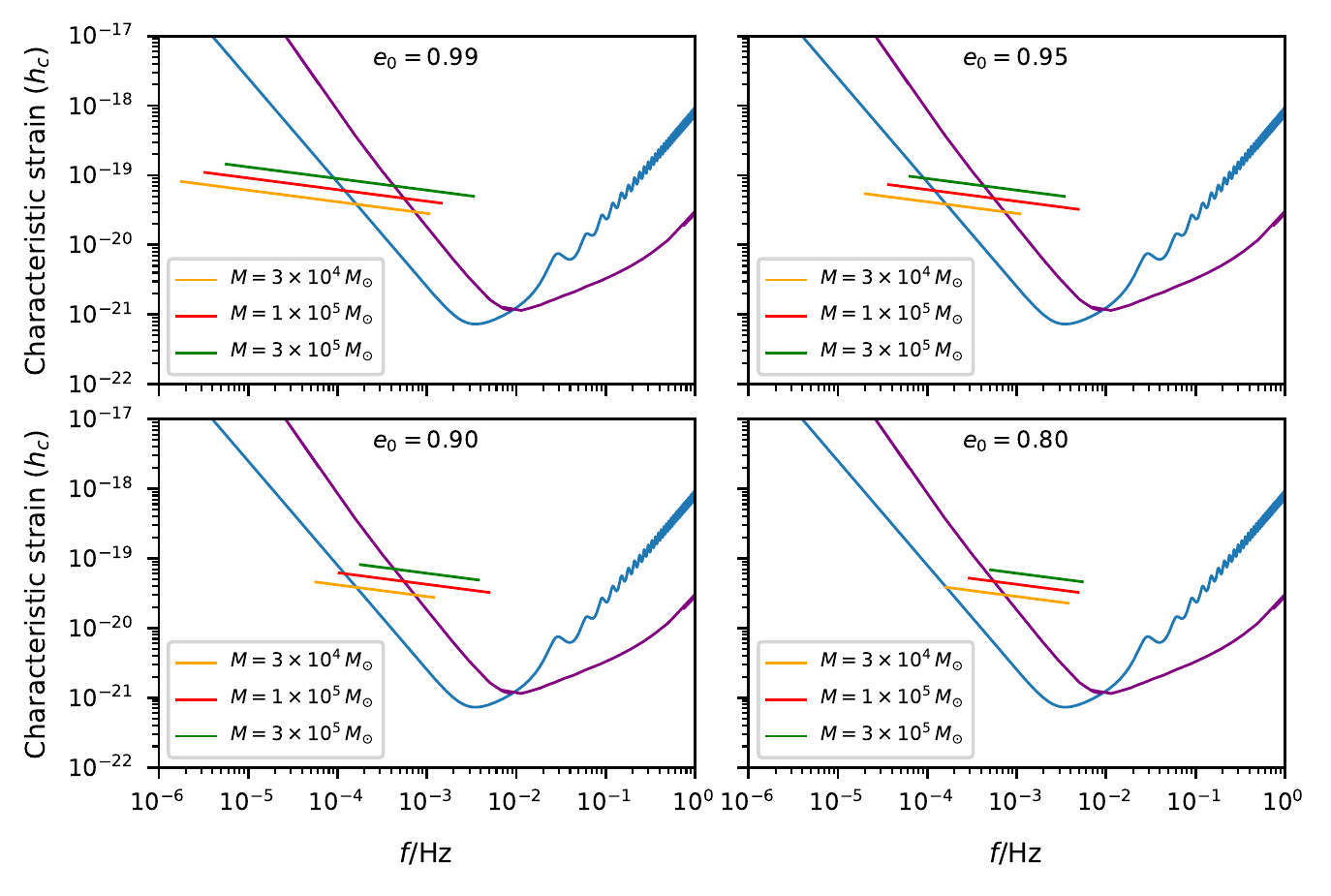}
    \caption{The diagram of EMRI's characteristic strain $h_c$ as a function of GW frequency $f$.
    For comparison, the sensitivity curves of LISA and Tianqin are plotted with blue and purple lines respectively.
    Each panel is depicted with a certain initial orbit eccentricity $e_0$ and different black hole masses.
    The mass of the C-O core is 3$M_\odot$ and the redshift of the EMRI system is $z = 0.2$ for all panels.}
    \label{Fig:Sensitivity curve}
\end{figure*}

The number of inspiral
cycles in the frequency range [$f_{\text{min}}, f_{\text{max}}$] is
given by
\begin{equation}
    N_{\text{cycles}} = \int_{f_{\text{min}}}^{f_{\text{max}}}\frac{f}{\dot{f}}\ud f.
\end{equation}
Typically, the small body will spend $10^{4-5}$ cycles inspiralling
into the SMBH, being observable for several years before plunge. The
characteristic strain $h_c$ of the GW from a source
emitting at frequency $f$ is
\citep{Finn00,Barack04,Maggiore18,Robson19,Amaro-Seoane18}
\begin{equation}
h_c(f) = 2f|\tilde{h}(f)| = \bigg(\frac{2f^2}{\dot{f}}\bigg)^{1/2}
h_0 = \frac{(2\dot{E}/\dot{f})^{1/2}}{\pi D},
\end{equation}
where $h_0$ is the instantaneous root-mean-square amplitude,
$\dot{E}$ is the GW emission power and $D$ is the proper distance to
the source. In our model, the characteristic strain $h_c$ is about
$10^{-19}$. It is worth mentioning that a fully coherent
search of $10^{4-5}$ cycles for EMRI detection is computationally
impossible. The feasible approach is hierarchical matched filtering
by dividing data into short data segments \citep{Gair04,Gair13}. The
signal-to-noise ratio (S/N) is built up in the second stage of the
search by incoherently adding the power of short segments
\citep{Gair04}, which will decreases by a factor $N^{-1/4}$ than a fully coherent search, where
$N$ is the number of divided segments \citep{Maggiore18}. An
incoherent search will be able to detect signals with $S/N \ge 20$;
while in a fully coherent search, the S/N required for detection is
12$\sim$14 \citep{Amaro-Seoane07,Babak17}. The S/N can be estimated
by \citep{Maggiore18,Robson19}
\begin{equation}
(S/N)^2 = \int_{f_1}^{f_2} \frac{h_c^2(f)}{h_n^2(f)}\ud (\ln f),
\end{equation}
where $h_n^2(f) = fS_n(f)$ and $S_n(f)$ is the noise spectral
density of the detector \citep{Maggiore18}. In our analysis,
a S/N threshold of 36 is assumed for incoherent search. Then EMRIs
formed in our channel can be detected as far as $z\sim 1$ (about 3.4
Gpc). The foreground noise from white dwarf binaries affects the detection of EMRIs, which has been
discussed by many authors \citep{Cornish03,Farmer03}. Some
algorithms are used to subtract this noise \citep{Cornish03} but
their performances are rather uncertain. However, even assume a 30\%
decrease of S/N after subtracting the WD background, the detection
range will not be less than $z \sim 0.7$. The schematic diagram of
EMRI's characteristic strain $h_c$ as a function of $f$ is shown in
Figure \ref{Fig:Sensitivity curve}. The LISA's sensitivity curve is
generated from the online sensitivity curve generator$\--$see
\cite{Larson03}.

By the way, the mass loss of the C-O core due to tidal
stripping after entering LISA band is less than 20\%, which may not
affect the detection of EMRI.

\section{Event rate}
\label{Sec:6}
In order to estimate the rate of
EMRIs occurring in the universe, two ingredients must be considered.
The first is the spatial density of SMBHs in the appropriate mass
range. The second is the rate at which each black hole tidally
disrupts massive stars. From observations, the space density of
SMBHs can be approximated by $M_{\rm BH}-\sigma$ relation
\begin{equation}
\label{msigma} M_{\rm BH}=M_{\rm
BH,*}\,\,\left(\frac{\sigma}{\sigma_{*}}\right)^{\lambda},
\end{equation}
where $\sigma$ is the spheroid velocity dispersion. We use
$\sigma_{*}=90\,{\rm km\,s^{-1}}$, $\lambda =4.72$ and $M_{\rm
BH,*}=3 \times 10^6 \,M_{\odot}$ \citep{Merritt01}. The above
relation is derived from SMBHs with masses ranging from
$10^6$ to $10^9$ $M_\odot$. For low-mass SMBHs
($<10^6M_\odot$), \cite{Xiao11} found that the $M_{\rm BH}-\sigma$
relation is consistent with the above relation allowing for
the uncertainties. Therefore, the $M_{\rm BH}-\sigma$ in
equation (\ref{msigma}) is used in our derivation. The galaxy
velocity dispersion function is constrained using galaxy luminosity
functions and $L\--\sigma$ correlation \citep{Aller02}. Combined
with the $M_{\rm BH}\--\sigma$ relation, the black hole mass
function is \citep{Gair04}
\begin{equation}
\label{bhdist} M_{\rm BH}\,\frac{\ud N}{\ud M_{\rm BH}} =
\phi_{*}\,\frac{\epsilon}{\Gamma\left(\frac{\gamma}{\epsilon}\right)}
\, \left(\frac{M_{\rm BH}}{M_{\rm BH,*}}\right)^{\gamma}\,
\exp\left[ -\left(\frac{M_{\rm BH}}{M_{\rm
BH,*}}\right)^{\epsilon}\right],
\end{equation}
where $\epsilon=3.08/\lambda$, $\phi_{*}$ is the total number
density of galaxies, and $\Gamma(z)$ is the Gamma function.
\cite{Aller02} derived the parameters $\phi_{*}$, $M_{\rm BH,*}$ and
$\gamma$ for different types of galaxies. For the mass
range of interest in this analysis, $M_{\rm BH} < 10^6 M_{\odot}$,
the parameters are $\phi_{*}=36.7~h_{70}^{2}\,{\rm Mpc}^{-3}$,
$M_{\rm BH,*}=4\times10^6M_{\odot}$ and $\gamma=0.03$. The spatial
density of black holes is approximately
\begin{equation}\label{BHM}
M_{\rm BH}\,\frac{\ud N}{\ud M_{\rm BH}} = 2\times 10^{-3}
\,h_{70}^{2}\,{\rm Mpc}^{-3},
\end{equation}
where $h_{70} \equiv H_{0}/70 \,{\rm km\,s^{-1}\,Mpc^{-1}}$ is the
dimensionless Hubble parameter.

The rate at which each SMBH disrupts massive stars can be calculated
using the loss cone theory \citep{Magorrian99,Wang04}. For solar-type
stars, the disruption rate per galaxy is \citep{Wang04}
\begin{equation}
\mathcal{R} =6.5\times10^{-4}\yr^{-1}
\left(\frac{M_*}{M_\odot}\right)^{-1/3}
\left(\frac{R_*}{R_\odot}\right)^{1/4}
\left(\frac{M_{\text{BH}}}{10^6M_\odot}\right)^{-1/4}.
\end{equation}
Using the standard Salpeter initial mass function, the number ratio
of 15-40$M_\odot$ stars to solar-type stars is $1.4\times10^{-2}$.
The lifetime ratio of massive star with 15$M_\odot$ to solar
type star is about $10^{-3}$. In addition, the typical density of the He envelope
is $10^3$ times larger than that of solar-type star, so the tidal
radius is one order of magnitude smaller. Hence, the rate should be
lowered by another factor of $10^{-3}$. What's more, for our
scenario to work, it is required that the star is on the He main
sequence, whose duration lasts roughly 0.1 times that of the H main
sequence. Combining all of the above factors and integrating
equation (\ref{BHM}) over $15\,M_{\odot}<M_*<40\,M_{\odot}$, $0.2
R_{\odot}< R_* < 8 R_{\odot}$, the event rate is $2\times
10^{-4}~\rm Gpc^{-3}yr^{-1}$ for $M_{\text{BH}}=5\times10^5M_\odot$.

Below, we briefly discuss how to identify this type of
EMRIs. From the spectrum of flare, the redshift of tidal
stripping event can be measured and the host
galaxy can be localized. After a few tens of years, LISA may detect
EMRI signal in the same direction, which will determine the
sky location to a few square degrees and the luminosity
distance to 10\% precision \citep{Babak17}. Combing the redshift
information from the flare with host galaxy properties, we can determine
whether the flare and the EMRI occur in the same galaxy.

\section{Summary}
\label{Sec:7} EMRI is a promising tool to study the strong
field gravity, the stellar dynamics in galactic nuclei, massive
black hole populations \citep{Babak17,Amaro-Seoane18} and many other
aspects of astrophysics. In this paper, we propose a new formation
channel for EMRIs, in which the tidal disruption flares can serve as 
EM precursor. The event rate of this type of EMRIs
is about $2\times 10^{-4}~\rm Gpc^{-3}\yr^{-1}$. Combined with relevant
EM signals, EMRIs will serve as a new standard siren to
probe the expansion of universe.

\acknowledgements We thank the anonymous referee for useful suggestions
which were helpful for improving the manuscript. We thank W.-B. Han, X. Chen and Ik Siong Heng for
helpful discussions. This work is supported by the National Natural
Science Foundation of China (grants U1831207, 11773010, U1738132,
1573014 and 11833003) and the National Key Research and Development
Program of China (grant No. 2017YFA0402600).

\end{document}